\def\defstyle#1#2{#1}

 \def\a{\alpha}\def\b{\beta}\def\g{\gamma}
  
  \def\th{\theta}\def\om{\omega}
 \def\G{\Gamma} 
 \def\m{\mu}  
 \def\P{\Psi} \def\ph{\phi}

\def\imo{i}

\def\be{\begin{equation}}
\def\ee{\end{equation}}
\def\bea{\begin{eqnarray}}
\def\eea{\end{eqnarray}}

\def\Re#1{\mathrm{Re}(#1)}
\def\Im#1{\mathrm{Im}(#1)}

\let\pd\partial
\def\mysection#1{{\bf #1}}

\defstyle{%
\documentclass[10pt]{article}
\def\auth#1{{\large #1}}
\def\affiliation#1{{\\\vspace{2mm}\it #1\\\vspace{5mm}}}
\def\mailbox#1{\footnote{Electronic address: #1}}
}{
\documentclass[twocolumn]{revtex4}
\let\auth\author
\let\mailbox\email
}%

\usepackage[dvips]{graphicx}
=-1cm
\begin{document}
\hyphenation{Schwar-zschild}

\title{High overtones of Dirac perturbations  of a Schwarzschild black hole}

\defstyle{%
\maketitle
\vspace{5mm}
\renewcommand{\thefootnote}{\fnsymbol{footnote}}
\begin{center}
}{%
}%
\auth{K. H. C. Castello-Branco}
 \mailbox{karlucio@fma.if.usp.br}
\affiliation{Universidade de S\~ao Paulo, Instituto de F\'\i sica
\\ Caixa Postal 66318, 05315-970, S\~ao Paulo-SP, Brazil.}

\auth{R. A. Konoplya}
 \mailbox{konoplya@fma.if.usp.br}
\affiliation{Universidade de S\~ao Paulo, Instituto de F\'\i sica
\\
Caixa Postal 66318, 05315-970, S\~ao Paulo-SP, Brazil.}

\auth{A. Zhidenko}
 \defstyle{\mailbox{Z\_A\_V@ukr.net}}{\mailbox{Z_A_V@ukr.net}}
\affiliation{Department of Physics, Dniepropetrovsk National University
\\
St. Naukova 13, Dniepropetrovsk  49050, Ukraine.}

\defstyle{%
\maketitle
\end{center}
\vspace{2mm}
\renewcommand{\thefootnote}{\arabic{footnote}}\setcounter{footnote}0
}{%
}%

\begin{abstract}
Using the Frobenius method, we find high overtones of the Dirac
quasinormal spectrum for the Schwarzschild black hole. At high
overtones, the spacing for imaginary part of $\omega_{n}$ is
equidistant and equals to $\Im{\omega_{n+1}}-\Im{\omega_{n}}
=i/8M$, ($M$ is the black hole mass), which is twice less than
that for fields of integer spin. At high overtones,  the real
part of $\omega_{n}$ goes to zero. This supports the suggestion
that the expected correspondence between quasinormal modes and
Barbero-Immirzi parameter in Loop Quantum Gravity is 
just a numerical coincidence.
\end{abstract}

\defstyle{%
\thispagestyle{empty}
\twocolumn
}{%
\maketitle
}%

The Quasinormal Mode (QNM) spectrum is an important characteristic
of a black hole. It dominates the late time response of a black
hole to an external perturbation, and, at the same time, does not
depend on the way of their  excitation. Thus, being dependent on
black hole parameters only, the QNMs provide us with the
"fingerprints" of a black hole, feasible to be seen in the
detection of gravitational waves (for reviews, see
\cite{kokkotas-review}).

The importance of the QN spectrum is not limited by the above
observational aspects of gravitational waves phenomena, since QNMs
have interpretation in conformal field theory through   AdS/CFT
and dS/CFT correspondence \cite{Birmingham2}. There is also
suggestion that the asymptotic  QNMs of black holes are connected
with the so-called  Barbero-Immirzi parameter in Loop Quantum
Gravity, which must be fixed to reproduce the Bekenstein-Hawking
entropy formula within this theory \cite{Dreyer}.
All this stimulated considerable interest to study the QNMs  of
black holes in flat, dS and AdS backgrounds \cite{kucha-mala}. 
In particular, the QNMs of the Dirac field for different black holes were considered in the
papers \cite{first-Cardoso}, \cite{Cho},
\cite{Zhidenko}, \cite{charged-dirac}. Nevertheless, the study of 
the Dirac QNMs of four-dimensional black holes  were limited 
by $low~ overtones$. The low overtones were found for a Schwarzschild black hole in
\cite{Cho}, with the help of the third-order WKB method
\cite{IyerWill}. There it
was observed that for low overtones the real part of the
frequency decreases as the damping grows. Soon, the analysis was
extended to the case of Schwarzschild-de Sitter black hole
\cite{Zhidenko} using the sixth-order WKB method
\cite{KonoplyaWKB6}, and also to the case of the charged Dirac
field in the background of a charged black hole
\cite{charged-dirac}.

In the present paper, we are trying to cover this gap in the study
of the Dirac QNMs and shall investigate the $high$ overtone
behavior of the Dirac perturbations of a Schwarzschild black hole.
We observed that at high overtones the imaginary  part of the QN
spectrum is equidistant with the spacing
$\Im{\omega_{n+1}}-\Im{\omega_{n}} =i/8M$. It is  also shown that
when the overtone number $n$ is increasing, the real part of
$\omega$ approaches zero.


The Dirac equation in an arbitrary curved background has the form
\cite{Birell}:
\be\label{Dirac}%
(\g^ae_a^{~\m}(\pd_\m+\G_\m)+m)\P=0,
\ee%
where $m$ is the mass of the Dirac field,  $e_a^{~\m}$ are
tetrads, $\G_\m$ are spin connections \cite{Birell}. The
Schwarzschild metric has the form:
\bea\label{metric}%
ds^2 = f(r)dt^2 - \frac{dr^2}{f(r)} - r^2 (d\th^2 + \sin^2\th
d\ph^2),
\eea%
where $f(r) =1 - (2 M/r)$, $M$ is the black hole mass. In this
background, the equation for the  massless Dirac field  can be
reduced, after some algebra, 
to the wave-like equation:
\be\label{Wave-like-equation}%
\left(\frac{d^2}{dr^{*2}} + \om^2 - V(r^*)\right)\Psi(r^*) = 0, \ee%
with the effective potential \cite{Wheeler}
\bea\label{potential}%
V(r)=f(r)\m\left(\frac{\m}{r^2}\pm\frac{d}{dr}\sqrt{\frac{f(r)}{r^2}}\right),
\eea%
which vanishes at the both boundaries: $V(r^*=\pm\infty)=0$. The
parameter $\m$ corresponds to a multipole number. Under the choice of the positive sign for the
real part of $\omega$ ($\omega =\omega_{Re} - i \omega_{Im}, \quad
\omega_{Re} > 0$),  QNMs satisfy the following boundary conditions

\be\label{bounds} \P(r^*) \sim C_\pm \exp(\pm\imo\om r^*), \qquad
r^{*}\longrightarrow \pm\infty,
\ee%
corresponding to purely in-going waves at the black hole event
horizon and purely out-going waves at infinity. Then, following
Leaver \cite{Leaver}, we can choose
\be\label{u-definition}%
\P(r^*)=\exp(\imo\om r^*)u(r^*),
\ee%
where $u(r)$ has a regular singularity at the  event horizon and
is finite at $r^*\longrightarrow\infty$.

The appropriate Frobenius series is
\be\label{Frobenius}%
u(r) = f(r)^{2s}\sum_{n=0}^\infty a_n f(r)^{n/2},
\ee%
\be f(r)^s\propto \exp(-\imo\om
r^*), s = -2M\imo\om, \quad r^*\longrightarrow-\infty.
\label{equation-s}
\ee
Substituting (\ref{u-definition}) and (\ref{Frobenius}) in
(\ref{Wave-like-equation}) we obtain the five-term recurrence
relation, which, after a series of Gaussian
eliminations can be reduced to the three-term recurrence relation

$$a_{n+1}\a_n(\omega)+a_n\b_n(\omega)+a_{n-1}\g_n(\omega)=0.$$

When $\omega$ is the quasinormal frequency, the ratio of the
series coefficients is finite and can be found from the standard
continued fractions \cite{Leaver}.


The QNMs are the roots of the inverted continued fraction:

$$\b_n-\frac{\a_{n-1}\g_{n}}{\b_{n-1}
-\frac{\a_{n-2}\g_{n-1}}{\b_{n-2}-\a_{n-3}\g_{n-2}/\ldots}}=
\frac{\a_n\g_{n+1}}{\b_{n+1}-\frac{\a_{n+1}\g_{n+2}}{\b_{n+2}-\a_{n+2}\g_{n+3}/\ldots}}$$
that can be solved numerically as soon as
$\a_n(\omega)$, $\b_n(\omega)$, $\g_n(\omega)$ are found.

Note, that for the above effective potential, we are not able to
use the Nollert expansion \cite{Nollert}. That is why in spite of
the slow convergence of the continued fractions in the unmodified
Leaver procedure we had to be limited by it. This certainly
requires much greater computing time to perform the computations.

\mysection{Low overtones.} The low overtones can be obtained either with the
help of the Frobenius method or, when the overtone number $n$ is
less than the multipole number $\mu$,  applying the WKB formula
\cite{IyerWill}, \cite{KonoplyaWKB6}.
\begin{table}
\centerline{$\mu=1$\hspace{3.5cm}$\mu=2$}
\centerline{\begin{tabular}{|c|c|c|}
  \hline
  $n$ & $\Re\omega$ & $-\Im\omega$ \\ \hline
\hline
  $0$ & $0.182963$ & $0.096982$\\ \hline
  $**$& $0.182639$ & $0.094938$\\ \hline
  $*$ & $0.176452$ & $0.100109$\\ \hline
  $1$ & $0.147822$ & $0.316928$\\ \hline
  $2$ & $0.146458$ & $0.424157$\\ \hline
  $3$ & $0.138885$ & $0.549184$\\ \hline
  $4$ & $0.134650$ & $0.674318$\\ \hline
  $5$ & $0.131726$ & $0.799895$\\ \hline
  $6$ & $0.128822$ & $0.925363$\\ \hline
  $7$ & $0.126064$ & $1.050710$\\ \hline
  $8$ & $0.123570$ & $1.176020$\\ \hline
\end{tabular}\quad
\begin{tabular}{|c|c|c|}
  \hline
  $n$ & $\Re\omega$ & $\-Im\omega$ \\ \hline
\hline
  $0$ & $0.380037$ & $0.096405$\\ \hline
  $**$& $0.380068$ & $0.096366$\\ \hline
  $*$ & $0.378627$ & $0.096542$\\ \hline
  $1$ & $0.355833$ & $0.297497$ \\ \hline
  $**$& $0.355857$ & $0.297271$\\ \hline
  $*$ & $0.353604$ & $0.298746$\\ \hline
  $2$ & $0.326095$ & $0.421677$ \\ \hline
  $3$ & $0.313180$ & $0.548007$ \\ \hline
  $4$ & $0.305740$ & $0.673165$ \\ \hline
  $5$ &  $0.300096$ &$0.798686$ \\ \hline
  $6$ &  $0.294725$ & $0.924309$\\ \hline
  $7$ & $0.289734$ & $1.049817$ \\ \hline
  $8$ &  $0.285201$ & $1.175236$\\ \hline
\end{tabular}
}%
\caption{First ten quasinormal frequencies of Dirac perturbations for
$\mu=1, 2$ found by the Frobenius method. For modes with $n <\mu$
the 6th order ($**$) and 3th order ($*$) WKB values are given.}
\end{table}
Let us remind that the potentials with opposite chirality produce
the same spectrum (see for instance  \cite{Zhidenko} and
references therein). That is why we shall consider only positive
values of $\mu$. The WKB approach has been used recently when
studying QNMs of different black holes (see \cite{WKB} and
references therein) and shown good agreement with numerical
computations for lower overtones. As can be seen from the tables
above we observe very good agreement between the WKB  and
Frobenius data, and, the higher the WKB order, the closer the
obtained WKB values to accurate numerical results.

\mysection{High overtones.}
The main difference from what we know on the high damping regime
for perturbations of fields of integer spin (scalar,
gravitational, and electromagnetic) is that now the spacing in
imaginary part is not $i/4 M$, as it takes place for integer spin
perturbations, but  $i/8 M$. The real part of $\omega$ (See
Fig.~\ref{fig1} and Fig.~\ref{fig2}) falls down quickly to tiny
values which already cannot be found with reasonable accuracy by
the Frobenius method. This occurs already at about $n=400-500$ for
$\mu=1$ multipole and at about $n =8000$ for $\mu=2$. Note that
the greater the ratio $Im\omega_{n}/Re\omega_{n}$, the more slowly
the Frobenius series converge, so we had to be careful when
finding high overtones and to check the convergence by increasing
of the ``length'' of the continued fraction until the result for
$\omega_{n}$ will not change. This required, for instance, the
length of the continued fraction up to 5 millions for $n = 7000$.
That is because we were not able to use the Nollert procedure for
the considered effective potential, and therefore, a lot of
computer time was required to get the convergence of the Leaver
procedure. That is the reason why we could not extend our
computation to higher $n$ than that shown on Fig~\ref{fig1}~and
Fig~\ref{fig2}. Note also, that for $n$ larger then $500$ it
is very time consumingly to compute the modes one by one, so
we had to ``skip''through the hundreds of overtones as it is shown on Fig~\ref{fig2}.  

Thus,  as can be seen from Fig.~\ref{fig1}~and~\ref{fig2}, QNMs
demonstrate the following asymptotic behavior:
\begin{equation}
\Re\omega_{n} \approx 0 \quad \textrm{as} \quad n \rightarrow \infty,
\quad
\end{equation}
\begin{equation}
\Im{\omega_{n+1}} - \Im{\omega_{n}} \approx -i/8 M \quad
\textrm{as}
\quad n\rightarrow\infty.
\end{equation}

\begin{figure*}
\vspace{-2cm}
\begin{center}
\centerline{\includegraphics{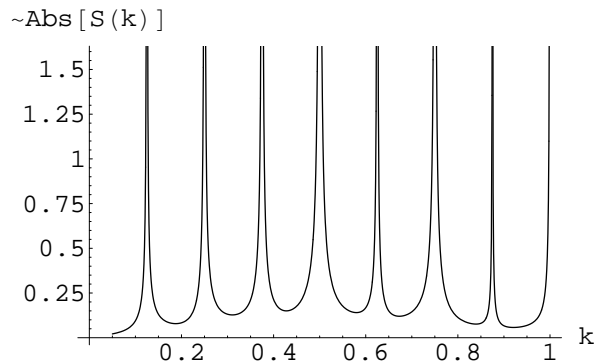}}
\caption{The absolute value of the scattering amplitude $S(k)$ as a
function of $k$ up to a constant factor. $M=1$, $\mu=1$.}
\label{fig0}
\end{center}
\end{figure*}

\begin{figure*}
\vspace{-2cm}
\begin{center}
\centerline{\includegraphics{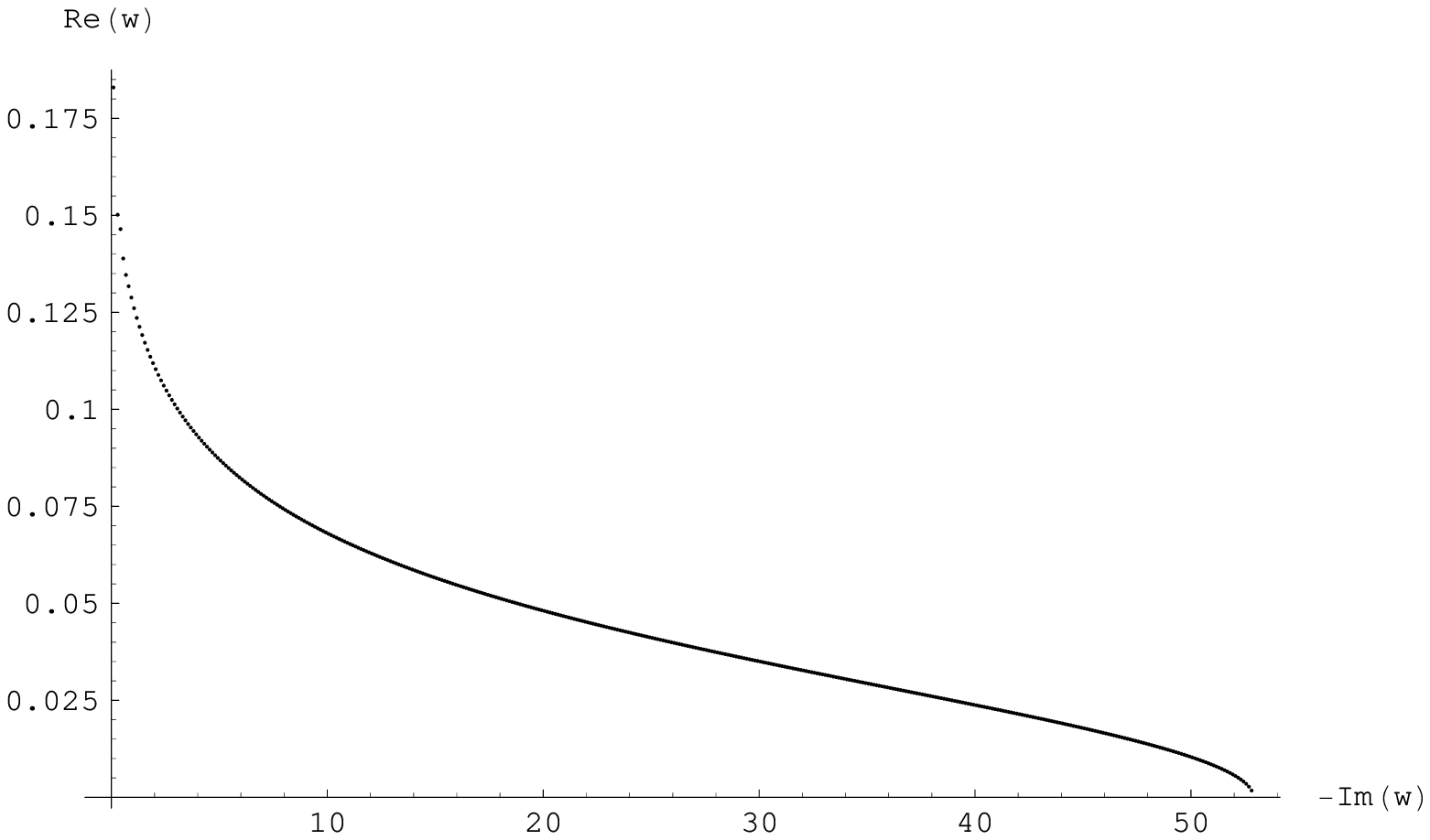}}
\caption{Real part of $\omega$ as a function of imaginary part for $\mu=1$
multi-pole.}
\label{fig1}
\end{center}
\end{figure*}

\begin{figure*}
\vspace{-2cm}
\begin{center}
\centerline{\includegraphics{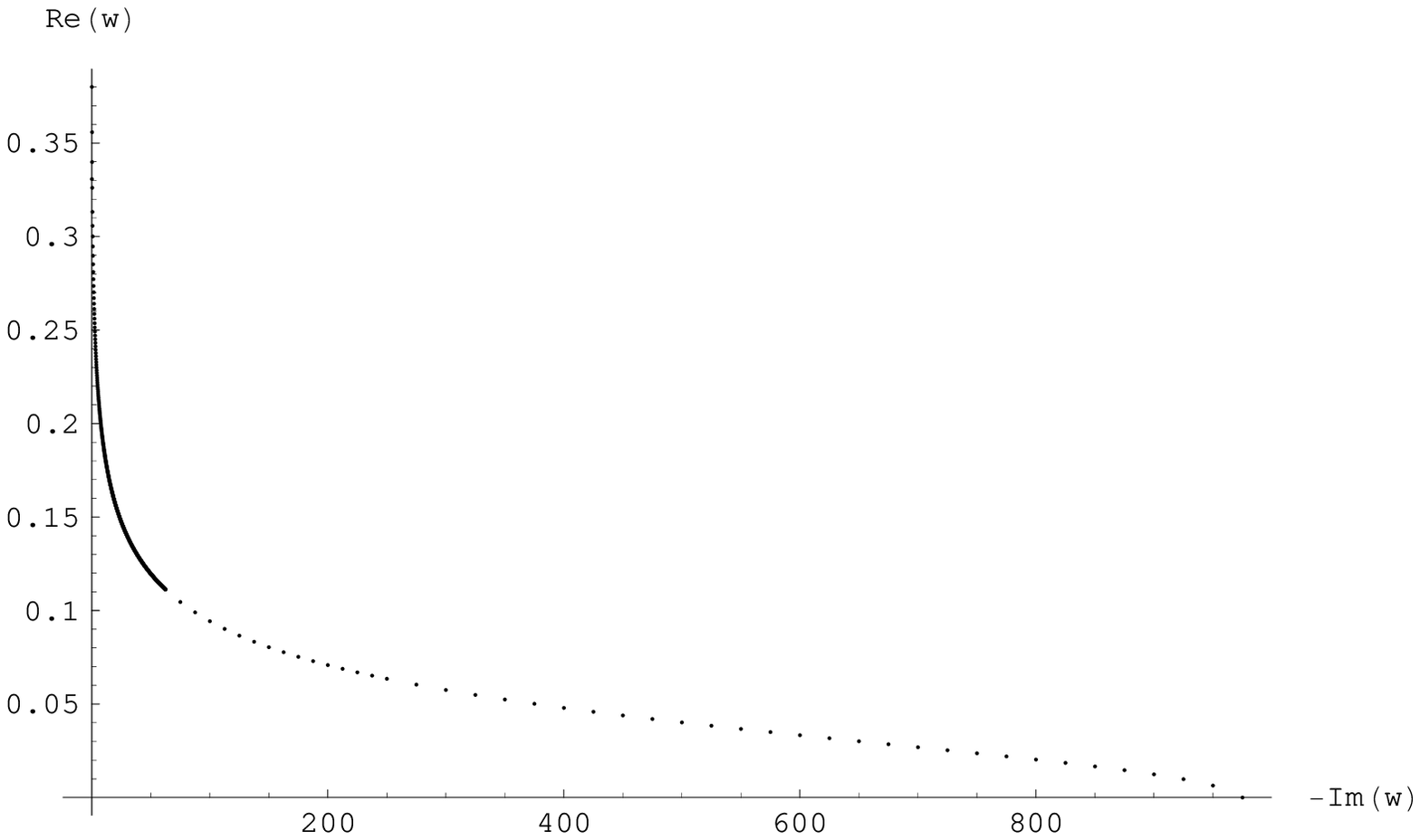}}
\caption{Real part of $\omega$ as a function of imaginary part for $\mu=2$
multi-pole.}
\label{fig2}
\end{center}
\end{figure*}

The formula for the spacing of the imagianary part can also be
reproduced following \cite{Pad}. Thus, for highly damping modes we
can use the Born approximation, where the scattering amplitude is
given by the formula \cite{Pad}:
\begin{equation}
S(k)= \int_{- \infty}^{+\infty} V[r(r^{*})] e^{2 i k r^{*}}dr^{*}.
\end{equation}
The above integral gets significant contribution only near the
event horizon. Using the Eq. \ref{potential} for $V[r(r^{*})]$, we find
$$ S(k) \sim~combinations~of~\Gamma (4 i k M)~ and~ \Gamma ((1/2) + 4 i k M ).$$
The poles of the amplitude occur when $(1/2) + 4 i M k =-n$, or $
4 i M k =-n$ ($n \geq 0$ and is integer), i.e. $k_{n} = i n/8 M$
is the required high frequency asymptotic for imaginary part. The
same result can be obtained either using the Taylor expansion of
the effective potential near the event horizon, or without such
expansion, but in the latter case the expression for the
scattering amplitude will have a cumbersome form. The
singularities of the scattering amplitude are shown on
Fig~\ref{fig0}.

\mysection{Conclusion.}
In this paper we have studied the high overtones of the Dirac
quasinormal spectrum for a Schwarz-schild black hole. The spacing
in imaginary part is two times less than that for integer spin
fields when $n \rightarrow \infty$. 
As was shown both numerically and analytically in \cite{referee2}
the real part of QNMs for scalar and gravitational perturbations 
asymptotically approaches  a constant equal to $\log 3/8 \pi$.
On the contrary, for the electromagnetic perturbations 
the real part goes to zero \cite{referee3}.
The real part for Dirac perturbations goes to zero, 
as it happens for electromagnetic
perturbations. This supports the suggestion that the expected
corresondence between quasinormal modes and  the Barbero-Immirzi
parameter in Loop Quantum Gravity is just a numerical coincidence
\cite{referee}.

The above analysis can easily be extended to the case of massive
Dirac field \cite{Cho}, \cite{Cho2}. At low overtones,  massive
Dirac pertubations \cite{Cho}, similar to massive scalar
pertubations  \cite{ya},  lead to greater oscillation frequencies 
and slower damping. At high overtones, similarly 
to the treatment in \cite{K-Zh} for massive scalar field, one 
can easily anticipate that the massive 
term will not affect the asymptotic quasinormal behavior.

\mysection{Acknowledgements.} The work of K. C-B. and R. K. was supported by
FAPESP (Brazil).

\end{document}